\documentclass[aps,prl,twocolumn,showpacs,superscriptaddress,floatfix]{revtex4}
\usepackage{epsfig}\usepackage{amsmath}
\usepackage{amsmath,amsfonts,amssymb,graphics,graphicx,epsfig,color,times,bbm}

\newcommand{\id}{{\sf 1 \hspace{-0.3ex} \rule{0.1ex}{1.52ex}\rule[-.01ex]{0.3ex}{0.1ex}}}

\begin{document}

\title{High efficiency transfer of quantum information and multi-particle entanglement
generation in translation invariant quantum chains.}

\author{Martin B. Plenio}

\affiliation{QOLS, Blackett Laboratory, Imperial College London,
Prince Consort Road, London SW7 2BW, UK}

\author{Fernando L Semi{\~a}o}
\affiliation{QOLS, Blackett Laboratory, Imperial College London,
Prince Consort Road, London SW7 2BW, UK}
\affiliation{Instituto de F\'\i sica Gleb Wataghin, Universidade
Estadual de Campinas, 13083-970 Campinas, S\~ao Paulo, Brazil}

\date{\today}

\begin{abstract}
We demonstrate that a translation invariant chain of interacting
quantum systems can be used for high efficiency transfer of
quantum entanglement and the generation of multi-particle
entanglement over large distances and between arbitrary sites
without the requirement of precise spatial or temporal control.
The scheme is largely insensitive to disorder and random coupling
strengths in the chain. We discuss harmonic oscillator systems
both in the case of arbitrary Gaussian states and in situations
when at most one excitation is in the system. The latter case
which we prove to be equivalent to an xy-spin chain may be used to
generate genuine multi particle entanglement. Such a 'quantum data
bus' may prove useful in future solid state architectures for
quantum information processing.
\end{abstract}


\maketitle

The realization of quantum communication and computation requires
at various stages the mapping between stationary and flying qubits
and the subsequent transfer of quantum information between
different units of our quantum information processing devices.
Traditionally the stationary forms of qubits are massive systems
such as atoms, ions, quantum dots or Josephson junctions while the
flying qubit is a photon, ie radiation. Photons might be optimal
when considering long distance communication where they may travel
through free space or optical fibres. In very small quantum
information processing devices such as condensed matter systems
however, this is difficult as the length scale both of the
component parts and their separation will generally be below
optical wavelengths. In this situation, it is worth considering
novel approaches for the communication of quantum information and
the generation of entanglement. To this end it is of interest to
consider the properties of interacting quantum systems and here in
particular those of harmonic systems that are realized in various
condensed matter physics settings such as nano-mechanical
oscillators. While static harmonic (or spin) systems near their
ground state do not exhibit long distance entanglement
\cite{Audenaert EPW 02}, the situation changes drastically when
considering time-dependent properties of interacting quantum
systems \cite{Khaneja G 02}. Indeed, solid state devices such as
arrays of nano-mechanical oscillators, described as interacting
harmonic oscillators, allow for the generation \cite{Eisert PBH
03}, transfer and manipulation of entanglement \cite{Plenio HE 04}
with a minimum of spatial and temporal control. However, in
translation invariant systems the efficiency for this transfer
decreased with distance. This can be overcome either by making the
coupling strengths between neighboring systems position dependent
\cite{Christandl DEL 03,Plenio HE 04} or by active steps such as
quantum repeater stages \cite{Osborne L 03} or conclusive transfer
\cite{Burgarth B 04}. Nevertheless, active steps or the
fabrication of precisely manufactured spatially dependent
couplings are difficult in practice and will require a significant
degree of control. Furthermore, the precise value of the coupling
parameters and the timing of the operations will depend on the
distance across which one aims to transfer quantum information.
Consequently, it would be desirable to achieve high efficiency
transmission of quantum information between arbitrary places and
distances with minimal spatial and temporal control. In the
following we show that this is indeed possible employing
translation invariant chains of interacting quantum systems with
stationary couplings.

We first describe the system, termed quantum data bus, and
demonstrate its functionality by numerical examples. Then we
present an approximate analytical model that reveals the basic
physical mechanism that is utilized in the operation of the
quantum data bus. This model then allows us maximize entanglement
transfer efficiency and transmission speed of the quantum data bus
by adjusting the eigenfrequencies of the sender and receiver
system. It also explains why the transmission is largely
insensitive to disorder and random coupling strengths in the ring.
We discuss the scaling behavior of the time that is required for
the transfer between distant sites at a given efficiency. Finally,
we show that in the regime where at most one excitation is in the
system, a quantum data bus made of interacting harmonic
oscillators becomes equivalent to an interacting spin chain. A
further application for such a chain is the generation of three or
multi-particle entangled states as we also show in this paper.
\begin{figure}[th]
\centerline{
\includegraphics[width=6.8cm]{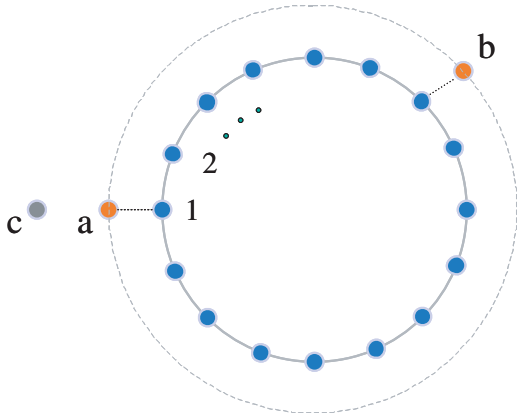}
} \caption{\label{fig} A ring of interacting quantum systems (blue
circles) forms the quantum data bus. Two further quantum systems
$a$ and $b$ (red) may couple weakly at arbitrary positions to the
ring. The subsequent time evolution will allow for the transfer of
quantum information or entanglement that exists between systems
$a$ and the decoupled system $c$ from system $a$ to system $b$.
System $c$ is decoupled but might be entangled with $a$ to study
transfer of entanglement. }
\end{figure}
The model we have in mind is depicted in Fig. 1. A ring of
interacting quantum systems (blue circles) forms the quantum data
bus. At arbitrary positions on the ring two further quantum
systems (red) may be coupled weakly to the ring. The subsequent
time evolution will allow for high efficiency transfer of
entanglement between the two distinguished quantum systems. The
following ideas are not restricted to the specific ring like
geometry presented here. The depicted geometry has been chosen
because it simplifies the analytical treatment of the problem.

So far we have not specified any particular type of quantum
systems nor their mutual interaction. In the following we will
consider as an example the case of coupled harmonic oscillators
which may be realized in various solid state settings \cite{Eisert
PBH 03} . Towards the end of this article we will also consider a
setting which is equivalent to a spin chain with an
xy-interaction. Let us assume that the ring consists of $M$
harmonically coupled identical oscillators. We set
$\hbar=\omega=m=1$ and denote the coupling strength in the ring by
$c$. We assume that oscillator $a$ ($b$) couples to oscillator $1$
($k$) with strength $\epsilon$ so that the Hamilton operator of
this system including the oscillators $a,b$ and $c$ is given by
\begin{eqnarray}
    H(\epsilon) &=& \frac{1}{2}\left[ \sum_{k=1}^M p_k^2 + \sum_{k,l=1}^M
    x_{k}V_{kl}x_{l} + \sum_{i=a,b,c} (x_i^2+p_i^2)\right]\nonumber\\
    && + \frac{\epsilon}{2} \left[ (x_a-x_1)^2 + (x_b-x_k)^2 \right]
    \label{Hamilton}
\end{eqnarray}
with the potential matrix $V$ given by $V_{kk}=1+2c$ and
$V_{k,k+1}=V_{k+1,k}=-c$ and $V_{1M}=V_{M1}=-c$ for all $k$ and
zero otherwise.

In the following we will demonstrate that it is indeed possible to
transmit quantum information through this system with high
efficiency but minimal spatial and temporal control. For the
moment we are focussing on Gaussian states, i.e., states whose
characteristic function or Wigner function is Gaussian. The
characteristic function determines the quantum state and as any
Gaussian is determined by its first an second moments the same
applies to the corresponding quantum state \cite{Eisert P 03}. In
the present setting the first moments will not be directly
relevant (they correspond to biases that can be removed by
redefining the coordinate origin). Therefore the state of the
system is determined by second moments that can be arranged in the
symmetric $2N\times 2N$-covariance matrix
    $\Gamma_{R,S}= 2\text{Re} \langle
    (R- \langle R\rangle )
    (S- \langle S\rangle )
    \rangle,$
where $R$ and $S$ stand for the canonical operators
$x_1,\ldots,x_n$ and $p_1,\ldots,p_n$.
%
\begin{figure}[th]
\centerline{
\includegraphics[width=8.0cm]{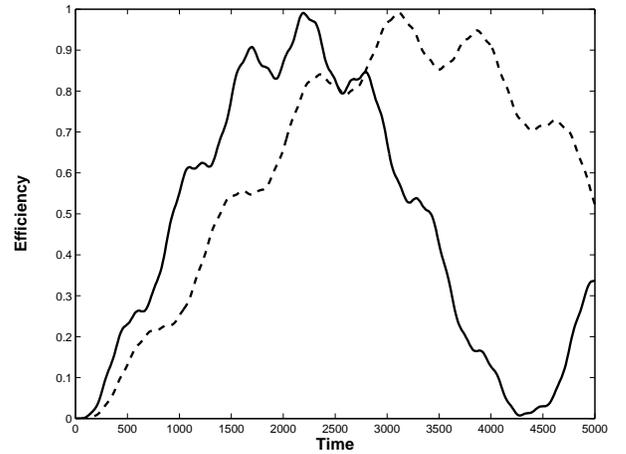}
}
\caption{\label{fig2} The efficiency of entanglement transfer,
defined as ratio of transmitted entanglement to initial
entanglement, between the oscillator $a$ and $b$ at opposite ends
of a quantum data bus consisting of $20$ oscillators and a nearest
neighbor coupling strength of $c=1$. The coupling strength of the
oscillators $a$ and $b$ to the quantum data bus is
$\epsilon=0.015$ (dashed line) and $\epsilon=0.021$ (solid line),
The speed of propagation is proportional to the coupling strength
$\epsilon$ and in both cases we observe maximal efficiency
exceeding $0.99$.}
\end{figure}
Employing the Hamilton operator eq. (\ref{Hamilton}) we can now
study numerically the quality of the entanglement transfer between
oscillators $a$ and $b$. Let us consider the situation where the
harmonic oscillator $a$ and $c$ are initially in a pure entangled
two-mode squeezed state
\begin{equation}
    |\psi\rangle = \sqrt{1-q^2}\sum_{n=1}^{\infty} q^n|n\rangle|n\rangle
\end{equation}
with $q=\tanh\frac{r}{2}$ which possesses the covariance matrix
\begin{eqnarray}
    \Gamma_{x_a x_c p_a p_c} = \frac{1}{2}
    \left(\begin{array}{cccc}
    \cosh r & 0 & -\sinh r & 0\\
    0 & \cosh r & 0 & \sinh r\\
    -\sinh r & 0 & \cosh r & 0\\
    0 & \sinh r & 0 & \cosh r
    \end{array}\right) \, .
\end{eqnarray}
The entanglement as quantified by the logarithmic negativity
\cite{Logneg} of this state is then given by
$E_N(|\psi\rangle)=2r$. The time evolution of the entanglement
between the oscillator $a$ and $b$ at opposite ends of a quantum
data bus consisting of $20$ oscillators and a nearest neighbor
coupling strength of $c=1$ is given in Fig. \ref{fig2}. The
propagation speed is proportional to $\epsilon$ and the efficiency
of entanglement transfer decreases weakly with increasing speed.
In both cases we find a maximal efficiency, defined as the ratio
of transmitted entanglement to initial entanglement, exceeding
$0.99$. It should be noted that the time required for the transfer
of entanglement is independent of the distance of places where the
oscillators $a$ and $b$ couple to the ring. The high quality of
the entanglement transfer and its independence on the position of
sender and receiver will be successfully explained by the
following model that encapsulates the essential physics in the
system.

We first observe that the unitary matrix $\Omega$ with matrix
elements
\begin{equation}
    \Omega_{kl} = \frac{1}{\sqrt{M}} e^{\frac{2\pi i k l}{M}}
\end{equation}

achieves $V=\Omega^{\dagger} \Lambda^2 \Omega$ with a diagonal
matrix $\Lambda$ such that
\begin{equation}
    \Lambda_{kk}^2 = 1+2c-2c\, cos\frac{2\pi k}{M} .
\end{equation}
Then we can define the normal mode variables
\begin{eqnarray}
    X_k &=& \sum_{l=1}^{M} \Omega_{kl} x_l, \;\;\;
    P_k = \sum_{l=1}^{M} \Omega_{kl}^* p_l
\end{eqnarray}
which ensure that $[X_k,P_l]=[x_k,p_l]=i\delta_{kl}$, ie the
canonical commutation relations are preserved. Note that we will
use the convention $X_0\equiv X_M$ and $P_0\equiv P_M$ which
reflects the periodic boundary conditions of the quantum data bus.
Further denote $X_{a,b,c}=x_{a,b,c}$ and $P_{a,b,c}=p_{a,b,c}$ to
make the notation more uniform. In these normal modes we can write
the Hamiltonian eq. (\ref{Hamilton}) as
\begin{eqnarray*}
    H(\epsilon) &=& \frac{1}{2} \left[\sum_{k=1}^M P^{\dagger}_{k} P_k
    + \Lambda_{kk}^2 X_{k}^{\dagger} X_{k} + \sum_{i=a,b} P_i^2 +
    (1+\epsilon) X_i^2
    \right]\\
    && + \frac{1}{2}(X_c^2 + P_c^2)
     - \epsilon X_a\sum_{l=1}^{M} \Omega_{1l}^*X_l
    - \epsilon X_b \sum_{l=1}^{M} \Omega_{kl}^*X_l\\
    && + \frac{\epsilon}{2} \sum_{lm}
    (\Omega_{l1}^*\Omega_{m1}^* + \Omega_{lk}^*\Omega_{mk}^*)
    X_lX_m\, .
\end{eqnarray*}
Defining the annihilation operators
\begin{eqnarray*}
    A_k = \frac{\Lambda_{kk}X_{k} + i P_{k}^{\dagger}}{\sqrt{2\Lambda_{kk}}}
    &&
    A_{a,b,c} = \frac{X_{a,b,c} + i P_{a,b,c}}{\sqrt{2}}
\end{eqnarray*}
we can rewrite the Hamilton operator in terms of the $A_k$.
Indeed, shifting the zero of energy and moving to an interaction
picture with respect to
\begin{eqnarray*}
    H = \frac{1}{2}\sum_{k=1}^M P^{\dagger}_{k} P_k
    + \Lambda_{kk}^2 X_{k}^{\dagger} X_{k} +
    \frac{1}{2}\sum_{i=a,b,c} P_i^2 + X_i^2
\end{eqnarray*}
we find
\begin{widetext}
\begin{eqnarray}
    H_I &=& \frac{\epsilon}{2} A_a^{\dagger}A_a
    + \frac{\epsilon}{2} A_b^{\dagger}A_b
    - \epsilon  \sum_{l=1}^{M}
    \frac{\Omega_{1l}^*(A_a e^{-it} + A_a^{\dagger} e^{it})+
    \Omega_{kl}^*(A_b e^{-it} + A_b^{\dagger} e^{it})}{2\sqrt{\Lambda_{ll}}}
    (A_l e^{-i\Lambda_{ll}t} + A_{M-l}^{\dagger}e^{i\Lambda_{ll}t})\nonumber\\
    &+&  \sum_{lm=1}^M
    \frac{\epsilon (\Omega_{l1}^*\Omega_{m1}^* + \Omega_{lk}^*\Omega_{mk}^*)
    (A_l e^{-i\Lambda_{ll}t} + A_{M-l}^{\dagger}e^{i\Lambda_{ll}t})
    (A_m e^{-i\Lambda_{mm}t} + A_{M-m}^{\dagger}e^{i\Lambda_{mm}t})}{2\sqrt{\Lambda_{ll}\Lambda_{mm}}}
    \, .
    \label{interaction}
\end{eqnarray}
\end{widetext}
In this interaction picture we find for $\epsilon=0$ that
$A^{I}_k(t)=A^{I}_k(0)$ while for finite $\epsilon$ we have
 $   \frac{d}{dt}A^I_k(t) = i[H_I,A^I_k(t)].$
In time dependent perturbation theory using
\begin{eqnarray*}
    \frac{A^I_k(t+\Delta t) - A^I_k(t)}{\Delta t} &\cong&  \frac{i}{\Delta t} \int_{t}^{t+\Delta t} ds_1
    [H_I(s_1),A^I_k(t)]
\end{eqnarray*}
and setting $\frac{1}{c} \ll \Delta t \ll \frac{1}{\epsilon}$ we
find that to first order in $\epsilon$ the modes described by
$A_a$ and $A_b$ are only coupled to one collective mode, namely
the center of mass mode, described by $A_M$. Therefore we can
ignore the contributions from all other eigenmodes. Shifting the
zero of energy again we finally obtain the simplified set of
equations
\begin{eqnarray}
    H_{approx} &=& \frac{\epsilon}{2} A_a^{\dagger}A_a + \frac{\epsilon}{2}
    A_b^{\dagger}A_b
    + \frac{2\epsilon}{M} A_M^{\dagger} A_M \nonumber \\
    &&\!\!\!\!\!\! - \frac{\epsilon}{2\sqrt{M}}
    [(A_a+A_b) A_M^{\dagger} + (A_a^{\dagger}+A_b^{\dagger})A_M ].
    \label{approximation}
\end{eqnarray}
Now we write this Hamiltonian again in the quadrature components.
Defining $P\equiv (P_c,P_a,P_b,P_M)^T$ and $X\equiv
(X_c,X_a,X_b,X_M)^T$ we find
\begin{eqnarray*}
    H_{approx} &=&
    P^TVP +
    X^TVX
\end{eqnarray*}
where
\begin{displaymath}
    V =\frac{1}{2}\left(\begin{array}{cccc}
    0 & 0 & 0 & 0\\
    0 & \frac{\epsilon}{2} & 0 &
    -\frac{\epsilon}{2\sqrt{M}}\\[0.1cm]
    0 & 0 & \frac{\epsilon}{2} &
    -\frac{\epsilon}{2\sqrt{M}}\\[0.1cm]
    0 & -\frac{\epsilon}{2\sqrt{M}} & -\frac{\epsilon}{2\sqrt{M}} & \frac{2\epsilon}{M}
    \end{array}\right).
\end{displaymath}
The above set of approximate equations of motion can be understood
intuitively by a simple mechanical model. Indeed they describe the
motion of a very heavy central pendulum (corresponding to the
oscillators in the quantum data bus) that is coupled weakly to two
comparatively light oscillators. From undergraduate mechanics we
know that if one of the light pendulum is originally oscillating
then, after some time, it will have stopped oscillating, while the
other light pendulum is now oscillating with almost the same
amplitude while the heavy central pendulum remains essentially at
rest. From this simple mechanical picture the dynamics in the
quantum setting that is described below can be understood quite
intuitively.

\begin{figure}[hbt]
\includegraphics[width=8.cm]{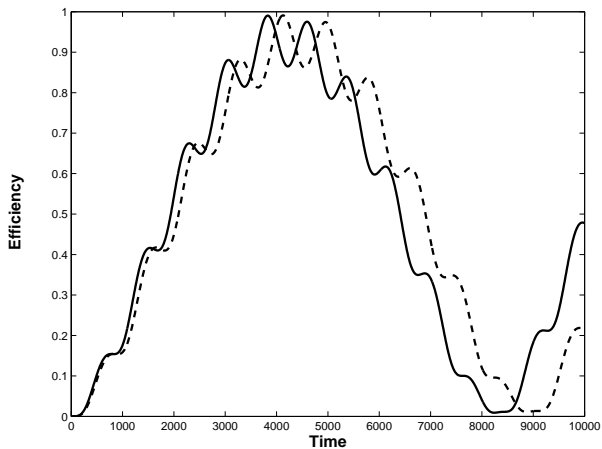}
\caption{\label{Fig3} Comparison of the approximation and the
exact time evolution for a ring of $20$ oscillators with $c=10$
and $\epsilon=0.021$. The dotted line is the approximation and the
solid line is exact.}
\end{figure}

That the above approximate Hamiltonian represents a good
approximation to the true dynamics can be seen from a comparison
of the exact time evolution with that generated by the approximate
Hamiltonian. In figure \ref{Fig3} we chose a ring with $20$
oscillators, $c=10$ and $\epsilon=0.021$. The observable mismatch
of between the frequencies is due to second order corrections to
the approximate model. Furthermore, as the excitation of a quantum
in the center of mass mode corresponds to the simultaneous
in-phase motion of all the oscillators in the quantum data bus and
the fact that the oscillators $a$ and $b$ couple predominantly to
the center of mass mode $M$ also explains why the entanglement
transfer is distance independent. The frequency of the
center-of-mass mode is independent of both the coupling strengths
between oscillators in the chain and possible disorder in which
oscillators are coupled as long as there is a connected path of
coupled oscillators between $a$ and $b$. Therefore the
transmission of quantum information between $a$ and $b$ will
depend only weakly on disorder and randomness. The main correction
arises when the frequency separation between the lowest two
eigen-modes becomes small so that off-resonant couplings of the
oscillators $a$ and $b$ to modes other than the center-of-mass
mode become non-negligible.
In the derivation of eq. (\ref{approximation}) we have neglected
many terms that led to oscillating contributions in the
Hamiltonian. These neglected terms will lead to corrections whose
size will depend on the length $M$ of the quantum data bus and
will affect both, the speed of propagation but also its
efficiency. As we can always adjust the waiting time, corrections
to the propagation speed are less relevant. An efficiency
reduction due to population losses is more serious as it will
require the application of error correction methods. For the
setting described by the Hamiltonian eq. (\ref{Hamilton}) we will
now present an estimate the size of these corrections. Indeed,
following eq. (\ref{interaction}) we neglect all terms that couple
the modes $a$ and $b$ non-resonantly to eigenmodes different than
the center of mass mode. For small coupling strength $\epsilon$
the rapid oscillations will reduce the population in these modes
significantly. The mode, other than the center of mass mode, with
the smallest oscillation frequency will be the mode $M-1$. For
large $M$ the frequency difference to the center of mass mode is
given by $\Delta=2\pi^2c/M^2$ and the coupling strength to this
mode is of the order of $\epsilon/2\sqrt{M}$. The loss of
population into this mode, which is of order
$\left(\frac{\epsilon\sqrt{M}}{2}\right)^2\left(\frac{4\pi^2c}{M^2}\right)^{-2}$,
should be small, ie
\begin{equation}
    \frac{\epsilon}{c} \ll \frac{4\pi^2}{M^{3/2}} \, .
\end{equation}
From eq. (\ref{approximation}) we determine that the speed of
propagation is proportional to $\epsilon/(2M)$. For a prescribed
transfer efficiency the transmission time therefore scales as
\begin{equation}
    T \sim \frac{2M}{\epsilon} \sim \frac{M^{5/2}}{2\pi^2 c} \, .
\end{equation}
The error source that enters this scaling is the population loss
to off-resonant modes. This suggest that this scaling can be
improved considerably when one allows for a fixed frequency
difference of the oscillators $a$ and $b$ compared to the
oscillators in the quantum data bus such that the oscillators $a$
and $b$ become resonant with a different collective mode whose
frequency difference to the neighboring modes is as large as
possible. Indeed, if we couple $a$ and $b$ to the mode described
by $A_{M/4}$, ie if we shift the eigenfrequencies of $a$ and $b$
by $2c$, then we find that the next mode is separated by a
frequency difference $\Delta=2c\pi/M$ and the couplings strength
to this neighbouring mode is again of the order of
$\epsilon/2\sqrt{M}$. As a consequence the population loss to
other modes is of the order of $(\epsilon\sqrt{M}/(4\pi c))^2$ and
we only need to ensure that
\begin{equation}
    \frac{\epsilon}{c} \ll \frac{4\pi}{\sqrt{M}}
\end{equation}
so that the propagation time for fixed transfer efficiency scales
as
\begin{equation}
    T \sim \frac{2M}{\epsilon} \sim \frac{M^{3/2}}{2\pi c} \, .
\end{equation}
This improved scaling has been achieved by coupling the
oscillators $a$ and $b$, but it should be noted that unlike the
center of mass mode $M$ this mode will have nodes. As a
consequence there will be relative positions of oscillator $a$ and
$b$ such that they will not couple, namely, when one is sitting at
a node while the other is at an anti-node. Indeed, the $M/4$ mode
has a node at every second oscillator of the ring, so that the
oscillators $a$ and $b$ couple only when their distance is an even
number of oscillators.

Finally we demonstrate that the above considerations are not
restricted to the continuous variable regime and the properties of
Gaussian continuous variable states. Indeed, we can equally well
consider a situation in which we restrict our dynamics to the
subspace spanned by the the vacuum and those states that
corresponds to a single excitation. The derivation of the
approximate model starting from Hamiltonian eq. (\ref{Hamilton})
remains of course valid. In the basis represented by the states
$\{|00\rangle_{ab}|0\rangle_M,|10\rangle_{ab}|0\rangle_M,
|01\rangle_{ab}|0\rangle_M,|00\rangle_{ab}|1\rangle_M\}$ where
$|1\rangle_M \equiv \frac{1}{\sqrt{M}} \sum_{r=1}^{M}
a_r^{\dagger}|vacuum\rangle$ we can rewrite the Hamiltonian eq.
(\ref{approximation}) as
\begin{eqnarray}
    H_{approx} &=& \frac{1}{2}\left(\frac{2\epsilon}{M}-\frac{\epsilon}{2}\right)
    (\id + \sigma_z^{M})\label{2spin}\\
    &&- \frac{\epsilon}{4\sqrt{M}}
    [(\sigma_x^a + \sigma_x^b)\sigma_x^M + (\sigma_y^a + \sigma_y^b)\sigma_y^M
    ]\nonumber\nonumber
\end{eqnarray}
which corresponds to a spin chain with an xy-interaction
\cite{Alves CJ} under the same constraint of considering at most a
single excitation. This similarity between a harmonic oscillator
systems and a spin chain is not due to the approximations in the
derivation of eq. (\ref{approximation}) but a generic feature when
one limits the number of excitations to at most one. A simple
computation shows that generally a harmonic chain in the rotating
wave-approximation and the single excitation regime will be
equivalent to a spin chain with xy-interaction in the same regime.

\begin{figure}[hbt]
\includegraphics[width=8.cm]{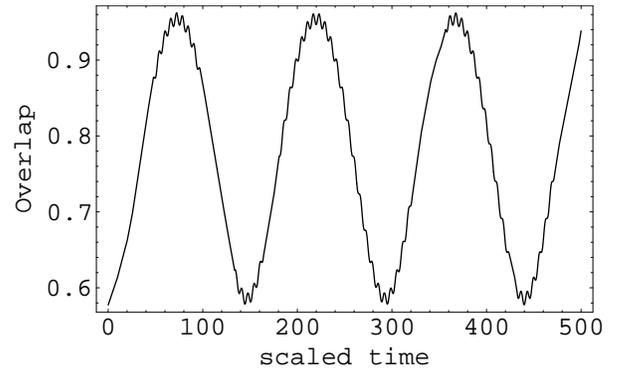}
\caption{\label{Fig4} Overlap between the W state with
$x=-y=-z=1/\sqrt{3}$ and the evolved state of the system. The ring
consists of $70$ oscillators.}
\end{figure}

\begin{figure}[hbt]
\includegraphics[width=8.cm]{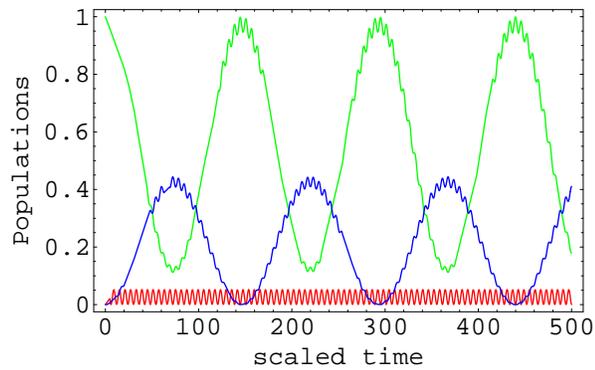}
\caption{\label{Fig5} Time evolution of the populations,
$|a(\tau)|^2$ in red, $|b(\tau)|^2$ in green, and
$|c(\tau)|^2=|d(\tau)|^2$ in blue. The ring consists of $70$
oscillators.}
\end{figure}

Continuing in this setting of a spin chain we now demonstrate that
multi-particle entanglement \cite{multiparticle} can be generated
with a quantum data bus extending the ideas employed in the paper
so far. While we focus on the discrete case, i.e. the spin
Hamiltonian, one may carry out a similar investigation for the
harmonic oscillator case. For the purpose of the generation of
entangled states there is no need of including the decoupled
oscillator in the discussion. While the ideas presented below may
easily be generalized to many oscillators let us, for simplicity,
consider three oscillators coupled to the chain. We assume that
the oscillator $a$ ($b$, $c$) couples to oscillator $1$ ($k_1$,
$k_2$) of the quantum data bus with strength $\epsilon$ so that
the Hamilton operator of this system is a generalization of eq.
(\ref{2spin}) and is given by
\begin{eqnarray}
    H &=&\frac{1}{2}\left(\frac{3\epsilon}{M}-\frac{\epsilon}{2}
    \right)(\id+\sigma_z^M)    \label{sp}\\
    && -\frac{\epsilon}{4
    \sqrt{M}}[(\sigma_x^a+\sigma_x^b+\sigma_x^c)\sigma_x^M +
    (\sigma_y^a+\sigma_y^b+\sigma_y^c)\sigma_y^M].
    \nonumber
\end{eqnarray}
Let us now suppose that the initial state of the system is
\begin{equation}
|\psi(0)\rangle=|1\rangle_a|0\rangle_b|0\rangle_c|0\rangle_M\equiv|1000\rangle
\end{equation}
where a single excitation is initially present in oscillator $a$.
The evolved state according to the Hamiltonian eq. (\ref{sp}) may
then be written as
\begin{equation}
    |\psi(t)\rangle=a(t)|0001\rangle+b(t)|1000\rangle+c(t)|0100\rangle+d(t)|0010\rangle
    \label{ex}
\end{equation}
where, with the scaled time $\tau\equiv \frac{\epsilon\, t}{2}$
and the constant $\alpha\equiv \frac{3-M}{M}$, as well as
\begin{eqnarray}
    C(M,\tau) &=& \cos(\sqrt{\frac{12+\alpha^2}{M}}\;\tau/2)\nonumber\\[0.2cm]
    S(M,\tau) &=& \sin(\sqrt{\frac{12+\alpha^2}{M}}\;\tau/2)
\end{eqnarray}
the time dependent coefficients are given by
\begin{eqnarray}
    a(\tau)&=& i \frac{e^{-i\alpha\tau/2}}{\sqrt{12+M\alpha^2}} C(M,\tau) \label{coef}\\[0.2cm]
    b(\tau)&=& \frac{2}{3}+ \frac{e^{-i\alpha\tau/2}}{3}
    \left[C(M,\tau) +  i\frac{S(M,\tau)}{\sqrt{12+M\alpha^2}}\right] \nonumber\\[0.2cm]
    c(\tau)&=& d(\tau)= -\frac{1}{3}+ \frac{e^{-i\alpha\tau/2}}{3}
    \left[C(M,\tau) + i\frac{S(M,\tau)}{\sqrt{12+M\alpha^2}}\right].\nonumber
\end{eqnarray}
From eq. (\ref{coef}) we can see that in the limit of large $M$,
the coefficient $a(\tau)$ tends to zero indicating that the
quantum data bus disentangles from the three oscillators and the
result may be a W state of the form
\begin{equation}
    |W_{x,y,z}\rangle\otimes|0\rangle_M
    =(x|100\rangle+y|010\rangle+z|001\rangle)\otimes|0\rangle_M
    \label{W}
\end{equation}
In order to see that this is the case we plot the overlap between
the W state eq. (\ref{W}) and the evolved state obtained with the
use of eq. (\ref{ex}) and eq. (\ref{coef}). For the case of $70$
oscillators in the chain, and the W state with
$x=-y=-z=1/\sqrt{3}$, it is shown in figure \ref{Fig4} that the
overlap is about 96\%. The overlap is not complete due to fact
that the population of the state $|0001\rangle$ does not vanish
and that the coefficients eq. (\ref{coef}) have a small imaginary
part. It should be noted that the second of these effects can be
reduced significantly by employing subsequent local unitary
rotations which are irrelevant when considering entanglement
properties of a state. In figure \ref{Fig5} we show the
populations in function of the scaled time $\tau$.
For generating different W states one could vary the coupling
constants independently what would lead to the following
Hamiltonian to the system
\begin{eqnarray}
H &=&
\frac{\epsilon_a}{4}(\id+\sigma_z^a)+\frac{\epsilon_b}{4}(\id+\sigma_z^b)
+\frac{\epsilon_c}{4}(\id+\sigma_z^c)\nonumber\\
&&+\frac{\epsilon_a+\epsilon_b+\epsilon_c}{2M}(\id+\sigma_z^M)
-\frac{\epsilon_a}{4\sqrt{M}}(\sigma_x^a\sigma_x^M+\sigma_y^a\sigma_y^M)\nonumber\\
&&-\frac{\epsilon_b}{4\sqrt{M}}(\sigma_x^b\sigma_x^M+\sigma_y^b\sigma_y^M)
-\frac{\epsilon_c}{4\sqrt{M}}(\sigma_x^c\sigma_x^M+\sigma_y^c\sigma_y^M).\nonumber
\end{eqnarray}
In summary, we have demonstrated that it is possible to transfer
quantum information with high efficiency but minimal spatial and
temporal control between arbitrary sites of a translation
invariant chain of quantum systems. We have shown that this
process works in the continuous variable regime but also the the
single excitation regime when the system becomes equivalent to the
dynamics exhibited by a single excitation in a spin chain with
xy-interaction. This interaction may be generalized to include
more oscillators coupled to the chain allowing the generation of
multi-particle entangled states. All these suggest that
translation invariant chains of interacting quantum systems are
promising candidates for the transport of quantum information in
solid state realizations of quantum information processing
devices.

We acknowledge discussions with S. Benjamin, S. Bose and A. Ekert
at an IRC discussion group and J. Eisert on many occasions. This
work is part of the QIP IRC (www.qipirc.org) supported by EPSRC
(GR/S82176/01) and by the Brazilian Agency Fundac{\~a}o de Amparo
a Pesquisa do Estado de S{\~a}o Paulo grant no 02/02715-2, and a
Royal Society Leverhulme Trust Senior Research Fellowship.

\end{document}